# Rotational Instabilities in Stratified Miscible Layers

**Saunak Sengupta[1], Sukhendu Ghosh[2], Sandeep Saha[3] and Suman Chakraborty[1]**

[1]Department of Mechanical Engineering, Indian Institute of Technology, Kharagpur-721302, India

[2]SRM Research Institute and Department of Mathematics, SRM Institute of Science and Technology, Chennai-603203, India

[3]Department of Aerospace Engineering, Indian Institute of Technology, Kharagpur-721302, India

## Abstract

Instability of stratified multi-phase flow in a rotating platform becomes important because of a potential role in micro-mixing and micro-machines. Centrifugal actuation can play an important role in driving the flow and Coriolis force can enhance the mixing in a short span by destabilizing the flow. In this study, we focus on the impact of the Coriolis force on a rotating viscosity-stratified flow with a thin diffusive mixed layer between two fluid layers. Modal stability analysis is used to estimate the critical parameters, namely Rotation number, Reynolds number, and wave number, which are responsible to modulate the instability mechanism for different viscosity contrasts. Present study explores competing influences of rotational forces against the viscous and inertial forces. Correspondingly, rotational direction (clockwise/anticlockwise) shows a significant effect on the spatio-temporal instability mechanism and anticlockwise rotation promotes the instability. Usually, miscible viscosity stratified flow with respect to streamwise disturbance becomes more unstable for a thinner mixed layer. On the contrary, our numerical computation confirms a completely contrasting scenario, considering Coriolis force driven instability of a miscible system on account of spanwise disturbances. Possible physical mechanisms for the same are discussed in terms of base flow and energy fluctuation among perturbed and base flow. Comparison of two and three-dimensional instability is done to give a clear-cut idea about the linear instability of the flow system considered herein. Velocity and viscosity perturbation distributions display a critical bonding between the vortices near and away from mixed layer, which may be responsible for the variation of instability with respect to viscosity ratio and rotational direction.

**Keywords:** Viscosity stratification, Miscible flow, Rotation, Coriolis force, Linear Stability Analysis

# 1. Introduction

Hydrodynamic instability of stratified (variation in density or viscosity or concentration etc.) multi-phase flow has drawn the attention of numerous researchers due to various industrial, chemical, biological and geophysical applications, which include lubrication, drag reduction, manufacturing of conjugated fibers, polymer melt, co-extrusion processes, filtration process, and several other flow of similar nature. In the recent years, bounded/semi-bounded flows with or without stratification in a rotating platform have been receiving progressively increasing attention because of its potential role in micro-mixing in portable medical diagnostic devices. Understanding instabilities in such flows are critical to several practical applications (Peter R.N. Childs, 2011), providing design and modelling capability for a diverse variety of products ranging from pumps, vacuum cleaners, jet engines etc on one side, and microfluidic mixer on the other.

In a wide variety of fluid flows, the viscosity is a function of space and time, resulting in viscosity stratification, which can have a startling effect on flow instability (Ghosh et al., 2014; Govindarajan and Sahu, 2014). Such a viscosity stratification can be achieved by various means, for example (i) considering two or more immiscible fluids which are in contact with each other by turns of a sharp interfacial discontinuity, (ii) including temperature or concentration difference between the fluids which can create a new thin diffusive layer, and (iii) deploying a non-Newtonian fluid where the effective viscosity is a non-linear function of space or imposed shear rate. In each of these cases, a disparity of fluid properties gives rise irregular interfaces, which leads to high scrap rates and undesired visual and mechanical properties (Khomami, 1990a, 1990b). As mentioned by Khomami (Khomami, 1990a, 1990b), to control various manufacturing and machinery operations better with regard to better design and improved efficiency, to select optimum processing conditions, and to choose different polymer systems with suitable rheological properties, a better understanding of multiphase flow processes, particularly with reference to the instability behaviour is very essential.

Earlier investigations on non-rotating viscosity stratified flow have often provided strikingly contradictory results concerning their possible roles on stabilization or destabilization. Yih (1963) was the first researcher probing the instability of multi-layer viscosity stratified flow. He focused on a long-wave perturbation analysis and arrived at the following significant conclusion: both plane Poiseuillle flow and plane Couette flow can be unstable for arbitrarily small Reynolds numbers due to viscosity stratification. Nevertheless, Loewenherz and Lawrence (1989) pointed out that multi-layer stratified free surface films can be unstable even in the inertia-less regime. Hickox (1971) applied Yih's approach later for an axisymmetric vertical two-fluid pipe flow with density stratification. Two important conclusions were obtained for the configuration with a less viscous fluid in the core of the pipe: first, the base/primary flow is always unstable to asymmetric or axisymmetric type disturbances, and second, the elementary reason for instability is the difference in viscosities of the two fluids. Thereafter, the multi-layer flow has been a topic of many other theoretical investigations (Ahrens et al., 1987; Charru and Fabre, 1994; Hooper and Boyd, 1983; Li, 1969; Malik and Hooper, 2005; Ranganathan and Govindarajan, 2001) and different instability mechanisms have been discussed. Based on the aforesaid studies and observations, it can be concluded that, to achieve a linearly stable flow, the less viscous fluid of thin layer must be placed closed to a wall to repose long waves and provide enough interfacial tension to overturn short waves.

Several researches have reported instabilities in stratified viscous layers considering variation either of temperature or concentration for flow inside a channel (South and Hooper, 1999; Wall and Wilson, 1996 Ghosh et al., 2014; Govindarajan, 2004; Malik and Hooper, 2005; Ranganathan and Govindarajan, 2001) or over an incline (Craik and Smith, 1968; Ghosh and Usha, 2016; Goyal et al., 2013; Hu et al., 2008; Kalliadasis et al., 2003; Usha et al., 2013). Correspondingly, Wall and Wilson (1996) explored the temperature dependent viscosity variation on a wall-bounded flow and showed that Peclet number has negligible influence on stability and the Tollmien-Schlichting (TS) type mode is subjugated by base viscosity variation. Investigation on inertia-less channel flow of elastic liquids (an Oldroyd-B fluid) having continuously stratified constitutive properties has beaconed instability due to the rapid normal stress variation (Wilson and Rallison, 1999). Various authors, notably Sahu and Matar (2010) and the cited authors in their study (Sukanek et al., 1973; Wyle and Huang, 2007; Yueh and Weng, 1996), considered the influence of viscous heating in different flows. They have shown a decrease in critical Reynolds number as the viscous heating increases. Moreover, extensive theoretical and experimental efforts have been invested to explore the salient features of the instabilities of interface dominant stratified fluid flows. Several factors, such as the properties of the fluids and the confining substrates, the viscosity and/or density stratification of the liquid layers, thickness ratio of the layers, jump in the velocities or the stresses across the interface and so on, are responsible for linear/nonlinear stability/instability of such interfaces (Goyal et al., 2013; Ó Náraigh et al., 2014).

Miscible multi-phase flows with confined geometry are special kind of stratified flows and reveal many interesting features of instability (Ern et al., 2003; Govindarajan, 2004; Malik and Hooper, 2005; Ranganathan and Govindarajan, 2001; Sahu et al., 2009). At very low miscibility, viscosity/density stratified flows showed similar instability characteristics as found in immiscible fluids. Investigations further suggested that a decrease in viscosity near the wall increases the critical Reynolds number enormously in some situations (Ranganathan and Govindarajan, 2001), but such a decrease can also cause the opposite: instability at extremely low Reynolds numbers (Selvam et al., 2007). Interfacial tension sometimes stabilizes the flow, sometimes destabilizes, and certain times does both. Following the work of Ranganathan and Govindarajan, (2001), Malik and Hooper, (2005) discussed linear stability and energy growth of a viscosity stratified flow after replacing liquid-liquid interface with a miscible layer of variable viscosity.

Ern et al., 2003) and Sahu et al. (2009) have discussed the influence of diffusion and mixed layer thickness on Couette type miscible two-fluid flow with a high degree of viscous stratification. They observed that the growth rate for the flow shows a non-monotonic behavior with respect to diffusion and such flows with intermediate Peclet number are much unstable for a thinner mixed layer. Govindarajan and co-workers (Govindarajan, 2004; Ranganathan and Govindarajan, 2001) deliberated the impact of a thin continuous stratified layer created by miscibility of two fluids with different viscosities and examined the payoff of slow diffusion. Quite interestingly, they identified a new mode of instability, which is very sensitive to the diffusion and associated with the overlapping of a critical layer and viscosity stratified layer. Thereafter, Talon and Meiburg (2011) considered stability analysis on similar kind channel flow in the Stokes flow regime and demonstrated that instabilities develop due to the diffusion between the layers. However, at large wave numbers, the onset of instability takes place at highly viscous region (in the core of the pipe). In the recent past, the work of Govindarajan (2004) was extended by Ghosh et al. (2014) to assess the effects of wall velocity slip on the instabilities of miscible viscosity stratified flow. It was showed that the flow system could either be stabilized or destabilized by designing the walls of the channel as slippery or hydrophobic surfaces, modelled by Navier-slip boundary condition.

Rotationally actuated micro-fluidic devices, like Lab-on-a-CD platform (Chakraborty et al., 2011) are inherently adept in generating secondary flows by virtue of the Coriolis force and outcome of this phenomenon has many advantages (Burger et al., 2012b, 2012a). Plane Poiseuille flow, subjected to a minor spanwise perturbation, becomes unstable at very low

Reynolds number (at $\leq O(100)$) in the presence of spanwise rotation (Alfredsson and Persson, 1989; Lezius and Johnston, 1976). The transitional rotating flow patterns were marked by the appearance of the streamwise roll-cell structure, which have been attributed to the Coriolis force (Chakraborty et al., 2011; Matsson and Alfredsson, 1990). Indeed, researchers have confirmed that the wavelength of the steady roll-cell instability can be predicted using linear stability theory. Intriguingly, the roll-call structures are gradually eliminated with an increase in rotational speed. Likewise, the experiments and the subsequent theoretical calculations (Chakraborty et al., 2011) on rotating flows reveal that the ratio of Coriolis force to the centrifugal force governs the extent of mixing achieved in such flows. The investigators reported that molecular diffusion dominates at low rotational speed due to the excess of centrifugal force rather than the Coriolis force. Mixing because of a streaky instability gets prominence when the rotational speed exceeds an emblematic critical value. The transverse Coriolis force is large enough to set up a secondary flow (Chakraborty et al., 2011) in such case. Numerical simulations of Roy et al. (2013) indicate that increasing the channel aspect ratio leads to a non-monotonic behaviour in the critical Reynolds number at which the secondary flow sets in, with the lowest corresponding to a square channel. Correspondingly, Wallin et al., (2013) mentioned that one can find the existence of oblique modes that have growth rates comparable to the streamwise-invariant mode (Wallin et al., 2013). Unstable modes sourcing from rotation may have much larger growth rate than that of the so-called two-dimensional spanwise-invariant Tollmien–Schlichting (TS) mode. These observations raise a question as to how the Coriolis force destabilizes the flow and if there is any passive technique to control the Coriolis force driven instabilities.

Stability issues governed by Coriolis force on viscosity stratified flows offer with many unanswered questions that can have a far-reaching consequence for future studies. For example, how varying the rotation rate affects the perturbation growth of a stratified flow, or how the fluid viscosity ratio influences the stability boundaries for the Coriolis-driven instability, remain unaddressed. Furthermore, how the presence of a miscible layer causes the mixing process in a rotating channel flow also remains unclear. Here, we bring out unique aspects of rotationally driven instabilities in stratified miscible viscous layers. One important aspect of the present investigation is to understand at what Reynolds number and rotation number the primary instability materializes in the subsistence of spanwise disturbances. In order to address the unanswered questions on viscosity stratified rotating channel flows and to progress towards a priory theoretical knowledge for designing guidelines of rotational flows, we conduct a modal stability analysis for the same.

The article is organized as follows: In section 2, we present the mathematical formulation of the problem and the set of governing equations in the rotating platform. In section 3, we put to address the stability regimes for infinitesimal disturbances (spanwise disturbances) with different viscosity ratios, mixed layer thicknesses, and clockwise/anticlockwise rotations. Influence of three-dimensional disturbances on the considered flow system is elaborated in section 4. Finally, a conclusion section contains the complete summary of the current study.

## 2. Theoretical formulation

The stability property of fluid flows having an interface with zero thickness is substantially different from those of the miscible flow. The difference is mainly because of the finite thickness mixed/interface layer, which occurs by slow diffusion and causes a smooth density and/or viscosity, concentration distribution instead of jump discontinuity. However, the thickness of the interface may not be constant and slowly vary with time. In the present study, we consider the modal linear stability of a laminar rotating three dimensional channel flow of two Newtonian, miscible, incompressible fluids. Study of this type of system is particularly appealing as it concerns a simple rotating shear type flow, which offers both stable or unstable regions depending on different ranges of parameters. Here, the two fluids are having the same density $\rho$, but different viscosities. A mixed layer with continuously varying viscosity separates the fluid layers with constant viscosity.

We emphasize that a rotational system with spanwise disturbance, where Coriolis force is required to drive the instability mechanism, molecular diffusion may play an important role on the instability mechanism for a flow with finite width interfacial region. We also remark that the parallel flow indisputably represents an idealization, as a real flow would have a diffuse interface whose thickness may vary in the streamwise distance. However, a parallel flow appears to be a good assumption and avoids the complication of analyzing instability behaviour for the miscible flows in the parameter regime we are working on. The flow is not necessarily symmetric about the centreline of the channel (see figure 1.). A Cartesian coordinate system is chosen for the present analysis, where $x$ and $y$ axes are respectively the mean flow and wall normal directions. The entire system is rotating with an angular velocity $\Omega$ about the $z$ axis whose direction follows the left-hand rule. The walls of the channel are located at the positions $y = \pm D$. Two fluids with viscosity $\mu_1$ and $\mu_2$ occupy the regions $-D \leq y \leq p$ and $p + q \leq y \leq +D$, respectively. The thin diffusive layer (between two constant viscosity fluid layers), referred as the mixed layer, occupies the region $p \leq y \leq p + q$, where two fluids are mixed up and a local smooth viscosity stratification is created.

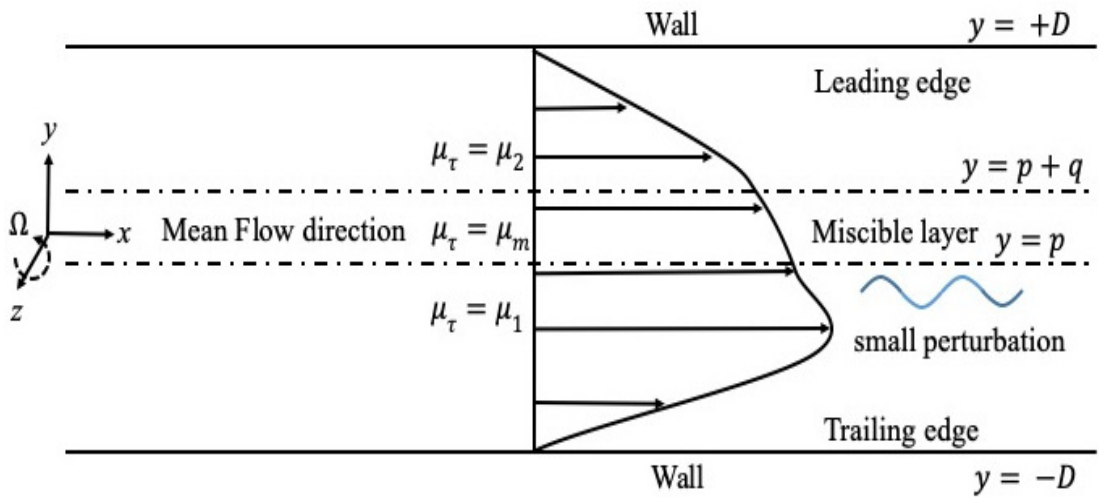

Figure 1. Schematic of the flow system considered. The entire system is rotating about *z*-axis, which is taken to be perpendicular to the plane of the paper.

## 2.1 Governing Equations

The governing differential equations for the mass and momentum conservation of the above rotating anatomy are (in dimensional from),

$$\nabla . \vec{u}_\tau = 0, \tag{1.1}$$

$$\frac{\partial \vec{u}_\tau}{\partial t} + (\vec{u}_\tau . \nabla) \vec{u}_\tau = -\frac{1}{\rho} \nabla p_\tau + \nu_\tau \nabla^2 \vec{u}_\tau - \frac{2}{\rho} (\vec{\Omega} \times \vec{u}_\tau), \tag{1.2}$$

$$p_\tau = \overline{p}_\tau - \rho \, \Omega^2 (x^2 + y^2), \tag{1.3}$$

where $\tau = u, m, l$ respectively pertain to the upper, mixed and the lower layer. The variables $\vec{u}_\tau$, $p_\tau$ and $\overline{p}_\tau$ are the velocity vector, modified pressure and pressure without rotation of the flow system, respectively. The parameter $\nu_\tau$ represents kinematic viscosity of each layer. The last term in the momentum equation originates from rotation ($\vec{\Omega} = \Omega \hat{k}$, $\hat{k}$ is the unit vector along $z$-axis) due to the presence of Coriolis force. The governing equations are non-dimensionalized with respect to the half channel width $D$, the maximum velocity of the flow $U_0$ and the viscosity of lower layer $\mu_1$. The dimensionless viscosity for each layers is given by $\mu_\tau(y)$ and

$$\mu_\tau\left(y\right) = \begin{cases} 1.0, & -1 \le y \le p \quad (\text{for } \tau = l) \\ \mu_m\left(y\right), & p \le y \le p+q \quad (\text{for } \tau = m) \\ r, & p+q \le y \le 1 \quad (\text{for } \tau = u) \end{cases}$$

where, $r = \mu_2 / \mu_1$ is the viscosity ratio of two fluids. The dimensionless viscosity $\mu_\tau(y)$ is chosen in such way so that the viscosity function and its first, second derivatives are continuous at the edge points of the miscible layer (i.e., at $y = p$ & $y = p+q$). This particular consideration can ensure the continuity in the linearized disturbance equation (Malik & Hooper 2005, Govindarajan 2004). Following the work of Malik & Hooper (2005), we have taken the viscosity of the miscible layer to be a fifth order polynomial with appropriate coefficients and given by,

$$\mu_m\left(y\right) = s_5 y^5 + s_4 y^4 + s_3 y^3 + s_2 y^2 + s_1 y + s_0 \tag{1.4}$$

where, the coefficients $s_0$, $s_1$, $s_2$, $s_3$, $s_4$ and $s_5$ are defined as follows ;

$$s_0 = 1 - \frac{p^3(r-1)}{q^5}(6p^2 + 15pq + 10q^2); s_1 = \frac{30p^2(r-1)}{q^5}(p^2 + 2pq + q^2); s_2 = -\frac{30p(r-1)}{q^5}(2p^2 + 3pq + q^2);$$

$$s_3 = \frac{10(r-1)}{q^5}(6p^2 + 6pq + q^2); s_4 = -\frac{15(r-1)}{q^5}(2p+q); s_5 = \frac{6(r-1)}{q^5}.$$

The unidirectional steady and fully developed basic flow is derived from the dimensionless equations using the no-slip and continuity conditions at the walls and interfaces, respectively. The dimensionless base velocity profile is identical to Malik & Hooper (2005) and read as,

$$U(y) = \begin{cases} \dfrac{1}{U_0}\left(-y^2 + (\delta-2)\,y + (\delta-1)\right), & -1 \le y \le p \\[2mm] \dfrac{1}{U_0}\displaystyle\int_p^y \left[\dfrac{-2\overline{y}+(\delta-2)}{\mu(\overline{y})}\right]d\overline{y} + U(p), & p \le y \le p+q \\[2mm] \dfrac{1}{U_0}\left(\dfrac{1}{r}\left[-y^2+(\delta-2)\,y+b_q\right]\right), & p+q \le y \le +1 \end{cases} \qquad (1.5)$$

where, the constant, $b_q = rU(p+q) + (p+q)[(p+q)-(\delta-2)]$. $U(y)$ satisfies zero-velocity at $y=-1$ and $U$, $\dfrac{dU}{dy}$ are continuous at $y=p$ and $y=p+q$. The coefficient $\delta$ is chosen appropriately in order to satisfy the no-slip condition at wall $y=1$.

## 2.2 Generalized Orr-Sommerfeld Square Equations

We focus on the linear stability analysis of the basic/mean flow with respect to three-dimensional small-amplitude disturbances in the form of $\varepsilon(y)\exp(i(\alpha x + \beta y - \omega t))$. Parameters $\alpha, \beta$ are the wave numbers of the disturbance in the stream and spanwise directions; $\varepsilon(y)$ and $\omega = \omega_r + i\omega_i$ are the amplitude and complex frequency, respectively. The complex phase speed $c$ of the perturbation wave satisfies $\omega = \alpha c$ and $\omega_i$ is the temporal growth or decay rate. We have used the normal mode formulation of normal velocity $(\tilde{v})$ and vorticity $(\tilde{\eta})$ perturbation given by $\{\tilde{v}, \tilde{\eta}\} = \{v, \eta\}\exp\left(i(\alpha x + \beta y - \omega t)\right)$ (Drazin and Reid, 2004) where,

$$\begin{Bmatrix} v \\ \eta \end{Bmatrix} = \begin{cases} \begin{Bmatrix} v_l \\ \eta_l \end{Bmatrix}, & -1 \le y \le p \\[3mm] \begin{Bmatrix} v_m \\ \eta_m \end{Bmatrix}, & p \le y \le p+q \\[3mm] \begin{Bmatrix} v_u \\ \eta_u \end{Bmatrix}, & p+q \le y \le +1 \end{cases}$$

defines the amplitudes of the velocity and vorticity disturbances at each layers. Perturbation quantities $\{\tilde{v}, \tilde{\eta}\}$ satisfy the generalized Orr-Sommerfeld Square equation for the variable viscosity (Govindarajan, 2004; Malik and Hooper, 2005).The final equations of disturbances in terms of wall normal velocity, vorticity yield,

$$-i\omega\left(v'' - k^2 v\right) + i\alpha U\left(v'' - k^2 v\right) - U''v = \frac{\mu_\tau}{Re}\left(v'''' - 2k^2 v'' + k^4 v\right) +$$
$$\frac{2\mu_\tau'}{Re}\left(v''' - k^2 v'\right) + \frac{2\mu_\tau''}{Re}\left(v'' + k^2 v\right) - i\beta Ro\,\eta \qquad (1.6)$$

$$-i\omega\eta + i\alpha U\eta + i\beta Uv = \frac{\mu_\tau}{Re}\left(\eta'' - k^2\eta\right) + \frac{1}{Re}\eta'\mu_\tau' + i\beta Ro\,v \qquad (1.7)$$

where, $Re \equiv \dfrac{\rho D U_0}{\mu_1}$, $Ro \equiv \dfrac{2D\Omega}{U_0}$ are the Reynolds and rotation numbers, respectively and prime ($'$) denotes differentiation with respect to $y$ (wall-normal direction). In the above equations $\mu_\tau$ is the base state viscosity. Moreover, $v$ and $\eta$ satisfy the no-slip and no-penetration boundary conditions at the channel walls and interface conditions at the edges of the mixed layer,

$$v(-1) = \eta(-1) = v'(-1) = 0 \tag{1.8}$$

$$v(+1) = \eta(+1) = v'(+1) = 0 \tag{1.9}$$

$$v_u(p+q) = v_m(p+q) \tag{1.10}$$

$$v_u'(p+q) = v_m'(p+q) \tag{1.11}$$

$$v_u''(p+q) = v_m''(p+q) \tag{1.12}$$

$$v_u'''(p+q) = v_m'''(p+q) \tag{1.13}$$

$$v_l(p) = v_m(p) \tag{1.14}$$

$$v_l'(p) = v_m'(p) \tag{1.15}$$

$$v_l''(p) = v_m''(p) \tag{1.16}$$

$$v_l'''(p) = v_m'''(p) \tag{1.17}$$

$$\eta_u(p+q) = \eta_m(p+q) \tag{1.18}$$

$$\eta_u'(p+q) = \eta_m'(p+q) \tag{1.19}$$

$$\eta_u'(p+q) = \eta_m'(p+q) \tag{1.20}$$

$$\eta_l(p) = \eta_m(p) \tag{1.21}$$

$$\eta_l'(p) = \eta_m'(p) \tag{1.22}$$

The above equations, together with all boundary conditions, are solved numerically using a spectral collocation scheme described in Schmid & Henningson (2001). The scheme and resulting numerical solutions are described in the next section.

## 3. Numerical Results and discussion

The generalized Orr-Sommerfeld Square equation along with the boundary conditions can be deduced to the following eigenvalue problem after numerical discretization:

$$A\Phi = -i\omega B\Phi,$$

where $A$ and $B$ are $9\times9$ block matrices and $\Phi = \left(\Phi_u \ \Phi_m \ \Phi_l\right)^T$ is the eigenfunction corresponding to the eigenvalue $\omega$. A hybrid Chebyshev spectral collocation method is used to solve the above eigenvalue problem. We have considered $nu+1, nm+1$ and $nl+1$ numbers of weighting functions in the upper, miscible and lower layer, respectively to approximate the eigenfunctions as follows

$$\Phi_u = \sum_{n=0}^{nu} a_n T_n(y), \ \ \Phi_m = \sum_{n=0}^{nm} b_n T_n(y), \ \ \Phi_l = \sum_{n=0}^{nl} c_n T_n(y),$$

where $T_n(y)$ represents the Chebyshev polynomial of first kind and $a_n$, $b_n$ and $c_n$ are unknown constants. Correspondingly, the Orr-Sommerfeld Square equation is implemented in each layer at the collocation points (Canuto et al., 1988):

$$y_j = \cos\left(\frac{\pi j}{M}\right), \qquad j = 1, 2, \ldots\ldots M-1,$$

$M = N-2$ with $N$ equals to $nu, nm$ and $nl$ points in the upper, middle and lower layers, respectively, giving $2(nu+nm+nl)-12$ equations. The rest comes from the boundary conditions and thus closing the system. We have used a developed MATLAB code to solve the above generalised eigenvalue problem. The accuracy of the numerical computation is achieved by varying the number of collocation points.

### 3.1 Parametric Results

We have elaborated different aspects of instability behaviour depending on various parameters with respect to small disturbances for the considered flow system. A wide range of new stability features of three-layer viscosity stratified miscible flow in a channel with spanwise system of rotation has been discussed. Those include the structure of eigen spectra, neutral stability boundary, eigen function behaviour and the role of rotational speed/direction on the flow system. We also have deliberated the effect of viscosity ratio and layer thickness as well. Eigen values are calculated numerically and limiting results are compared with the available results for the non-rotating single-fluid flow (Orszag, 1971; Schmid and Henningson, 2001) alongside the rotating single fluid system investigated by Alfredsson & Persson (1989).

Validation of our numerical code is first addressed in figure 2(a) by comparing our results with that of Orszag (1971), Drazin & Reid (2004) and Schmid & Henningson (2001) for the single fluid Poiseuille flow when $Re = 10,000$ and $\alpha = 1.0$, in absence of viscosity contrast $(r \to 1.0)$. We have obtained the classical Y-shaped eigen-spectrum with three principal branches ($A$, $P$, $S$) and a portion of discrete eigenvalues for nonzero $\alpha$ and zero $\beta$ is shown. The associated most unstable shear mode is located on '$A$' branch and corresponds the value $\omega = 0.237526 + 0.0037396i$. Further, we have computed the neutral stability boundary from the generalized eigenvalue problem with $\omega = \omega(\alpha, \beta, Re, Ro, r, p, q)$ for the rotating flow when viscosity ratio $r \to 1.0$ and instability is determined by the condition (imaginary part of the eigenvalue) $\omega_i > 0$. Neutral stability is achieved by satisfying, $\omega_i = 0$. Note that, the case $r \to 1.0$ corresponds to a single fluid system. We have compared our single fluid limiting results with Alfredsson & Persson (1989) in figure 2(b) and have found a very good agreement.

In figure 3, we have demonstrated the stability boundaries in $\beta - Re$ plane at different rotation numbers and viscosity ratios for both anticlockwise and clockwise rotation, without considering streamwise disturbances. In figures 3(b) and 3(d), the direction of rotation is clockwise and it is other way for figures 3(a) and (c). Notably, clockwise (CW) and anticlockwise (ACW) rotations, respectively, implicate negative and positive rotational number. Neutral stability diagrams show distinguishable effects of CW, ACW rotations and in the case of ACW it is more unstable. In the case of rotating single fluid Lezius & Johnston reported that the critical disturbance mode occurs at the critical Reynolds number $Re_{cr} = 88.53$ and the rotation number $Ro = 0.5$. Furthermore, increase in the rotation number stabilizes the system via increasing the $Re_{cr}$ value. In congruence with Lezius & Johnston (1976), we have found the similar phenomenon for the viscosity stratified rotating channel flow, and both the critical Reynolds number ($Re_{cr}$), critical rotation number ($Ro_{cr}$) vary with respect to the viscosity contrast. However, scaling of their study is different from the current study. Consequently, the value of $Re_{cr}$ and $Ro_{cr}$ alter as we change the direction of the rotation (CW or ACW). For a small viscosity variation with $r = 1.2$, our computed critical value of the rotation number ($Ro_{cr}$), Reynolds number ($Re_{cr}$) and spanwise wavenumber ($\beta_{cr}$) for anticlockwise rotation are $Ro_{cr} = 0.35, Re_{cr} = 75.38$ and $\beta_{cr} = 2.56$, respectively.

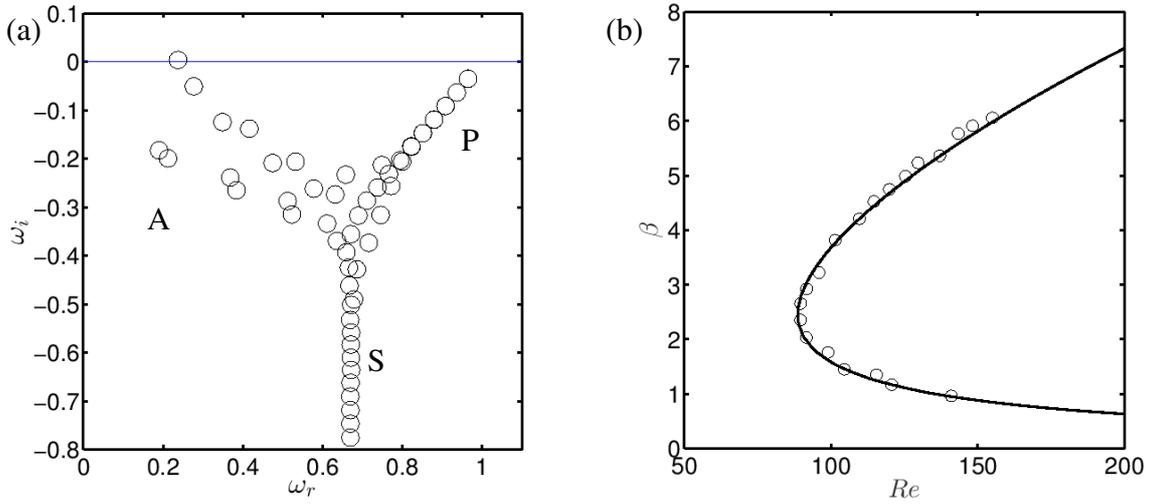

Figure 2. (a) Orr-Sommerfeld spectrum for non-rotating single-fluid limit ($r \to 1.0, Ro = 0$) from the two-fluid miscible viscosity stratified Poiseuille type flow when $Re = 10,000$ and wave numbers $\alpha = 1, \beta = 0$. (b) Neutral stability boundary for rotating channel flow applying single-fluid limit with rotation number $Ro = 0.5$. Solid line presents our computation result and circles are from Alfredsson & Persson (1989) result.

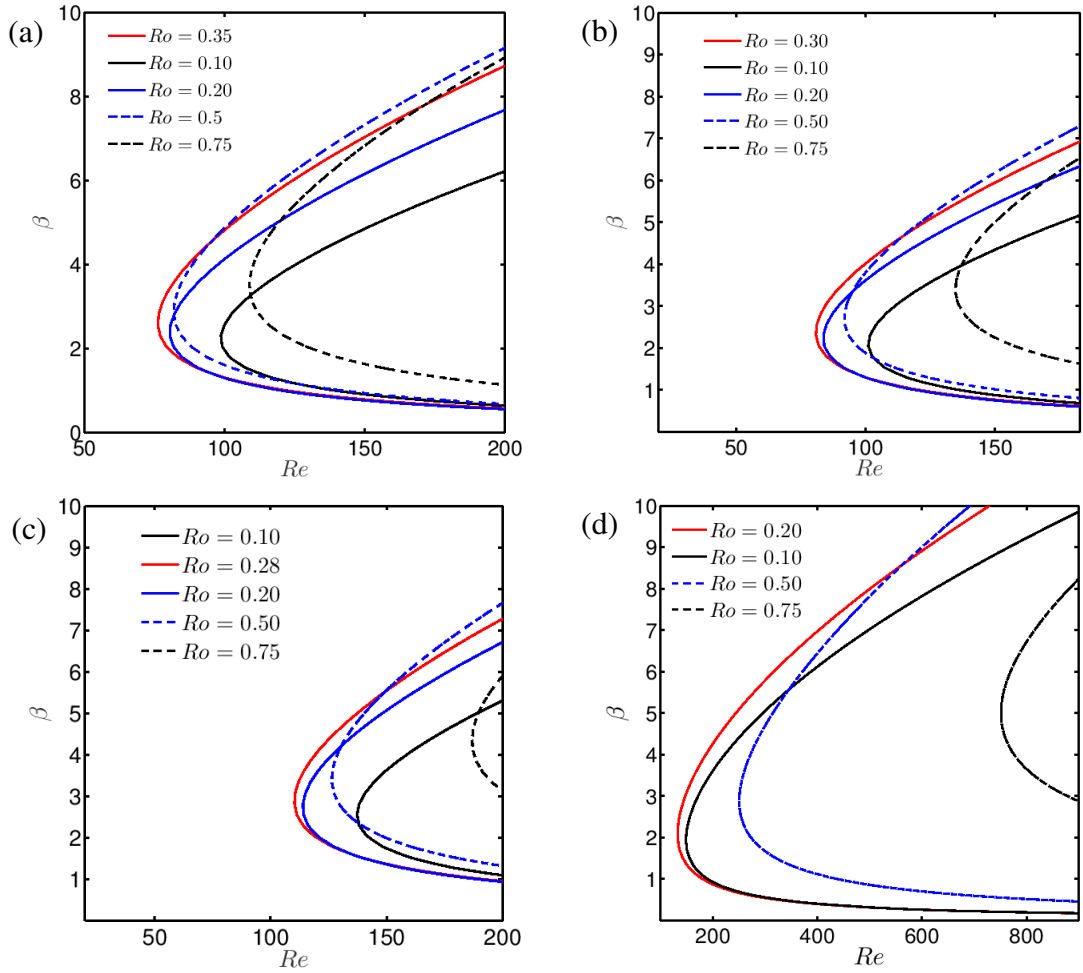

Figure 3. Neutral stability boundaries for two different viscosity ratios $(r)$: in (a) and (b) panels $r = 1.2$ and in (c) and (d) panels $r = 2.0$. Right and left columns of the figure are for clockwise and anticlockwise rotation, respectively.

The relevant parameters are $Ro_{cr} = 0.3$, $Re_{cr} = 80.023$ and $\beta_{cr} = 2.3$ for clockwise rotation. Further, increasing the viscosity ratio to $r = 2.0$, a similar trend is observed in the variation of critical values (we found $Re_{cr}$ increases to $112.31$ but $Ro_{cr}$ reduces to $0.28$, for anticlockwise rotation). Inclusion of the viscosity variation causes an asymmetry in the base flow (for $r = 1.2$ gradient of the base flow velocity is higher than that for $r = 2.0$), coupled with the rotational effect, renders the system more susceptible to rotation-induced instability. It is to be noted that for a symmetrical parabolic profile, reversing the direction of rotation does not affect the neutral stability boundary. For the case of symmetric base velocity profile where the direction of rotation does not affect the neutral stability boundary may be because of the fact that the absolute value of wall normal pressure gradient from central line is same for both the ACW and CW rotation. However, with the viscosity contrast, if a fluid element is considered at the same distance above and below the central line, absolute value of the wall normal pressure gradient no longer remains to be the same like the case of single fluid flow. Henceforth, $Ro_{cr}$ adopts lower value for higher viscosity deference between upper and lower layer. From figure 3, we can infer that beyond $Ro_{cr}$ value, the flow becomes unstable at much higher $Re$ if we continue to increase the rotational speed, which confirms the stabilizing mechanism of large rotation number as discussed by Tritton & Davies (1985).

The parameter $p$ represents the location around which the miscible layer starts and $q$ denotes the mixed layer thickness. Both the parameters $p$ and $q$ are very important for this particular type of flow, as they determine the relative mass fluxes of the fluid layers and the skewness of the base velocity profile (other two critical parameters are the direction of rotation and viscosity ratio). Thus, we aim to examine the influence of mixed layer thickness on the flow instability for both ACW and CW rotation. The neutral stability boundaries for two different viscosity ratios $r = 1.2$ and $2.0$ are shown in figure 4 for both CW and ACW rotations with variant mixed layer thicknesses. For instance, two different values of mixed layer thickness are chosen to understand the effect of viscosity stratification and rotational direction on the stability boundaries for a thinner $(q = 0.1)$ and a thicker $(q = 0.3)$ mixed layer. Earlier investigations on miscible non-rotating flows (Ghosh et al., 2014; Govindarajan, 2004; Malik and Hooper, 2005) indicate that the flow become more unstable with respect to streamwise disturbances when thickness of miscible layer is less. In the inset of the figure 4(a), we have showed how our code captures the physics of a two-dimensional non-rotating flow for viscosity ratio $r = 1.2$, which is consistent with the earlier reported results. It is significant and interesting to note that results for the viscosity-stratified flow with a spanwise system of rotation unveil contrasting physics as compared to what is known earlier in the absence of any kind of rotation. Notably, the present flow system with rotation is less unstable for the case of thinner mixed layer, which is the most important and extra special finding of the current investigation.

One may observe from figure 4(a) that the effect of rotational direction on the neutral stability boundary in the $\beta - Re$ plane is not very promising for $r = 1.2$, even though we varied the mixed layer thickness $(q)$. However, for quite thinner mixed layer $(q = 0.1)$, the influence of rotational direction is much significant and the flow tends to become more stable at $\beta - Re$ plane with the aid of viscous forces (although $Ro_{cr}$ is lesser in the case of CW rotation). The critical Reynolds number remains almost invariant with the variation of layer thickness for ACW rotation when viscosity ratio $r = 1.2$. The reason behind this phenomenon is that, the velocity profile does not get affected much by low viscosity contrast. Enhancing the viscosity ratio to $r = 2.0$ (figure 4(b)), we see a higher order change in the value of $Re_{cr}$ if we switch the mixed layer thickness. This may be owing to the excess of maximum base velocity for mixed layer thickness $q = 0.3$ than that corresponding to $q = 0.1$ (to be noted, we have defined our Reynolds number based on maximum velocity of the channel). Nevertheless, for clockwise rotation, decreasing the miscible layer thickness makes the flow more stable. Quite interestingly, the critical value of Reynolds number is achieved at a higher value with an increase of viscosity ratio and it turns out to be a more stable configuration with higher viscous fluid is adjacent to the upper wall. In Table-1, we have enlisted the critical value of rotation number, Reynolds number and spanwise wavenumber for three different viscosity ratios $r = 0.8$, $1.2$ and $2.0$ for both clockwise rotation and anticlockwise rotation. It is worth mentioning that, $r = 1.2$, $2.0$ suggest the configuration where a fluid with higher viscosity flows over a low viscous fluid and $r = 0.8$ yields an opposite flow configuration. From the table it is clear that disturbances are unstable at lower $\beta_{cr}$, $Ro_{cr}$ for anticlockwise rotation when $r < 1$ and the same is true for clockwise rotation when $r > 1$.

| $r$ | $\beta_{cr}$ | $Ro_{cr}$ | $Re_{cr}$ |
|-----|--------------|-----------|-----------|
| 0.8 | 2.41 | $0.33(\text{ACW})$ | 56.90 |
| 0.8 | 2.76 | $0.40(\text{CW})$ | 53.44 |
| 1.2 | 2.56 | $0.35(\text{ACW})$ | 76.38 |
| 1.2 | 2.3 | $0.30(\text{CW})$ | 81.02 |
| 2.0 | 2.82 | $0.28(\text{ACW})$ | 112.31 |
| 2.0 | 2.034 | $0.20(\text{CW})$ | 136.44 |

Table I. Critical values of the spanwise wave number $(\beta_{cr})$, rotation number $(Ro_{cr})$ and the Reynolds number $(Re_{cr})$. A clockwise rotation signifies negative rotation number but the sign is omitted here.

To give additional physical insight about evolution of flow instability, the wave amplitude pattern of the wall normal velocity and vorticity (which are nothing but the eigen functions) need to be comprehensively assessed. The real, imaginary and the absolute value of the typical eigen functions for most excited eigen mode are assembled in the figures 5 with both CW and ACW rotations. All the eigen functions are normalized using supreme norm to make the comparison effective. Disturbances of the spanwise and streamwise velocity components are proportional to the normal vorticity and derivative of normal velocity components, respectively. Figures 5(a), (b) and 5(c), (d) are drawn for the case when the more viscous fluid is adjacent to the lower $(r = 0.8)$ wall and upper $(r = 1.2)$ wall, respectively. The critical perturbation of the wall normal velocity attends a maximum value in the bulk of fluid (having less/high viscosity for $r = 1.2/0.8$) adjacent to the lower wall for anticlockwise rotation and in the fluid region near to the upper wall (having high/less viscosity for $r = 1.2/0.8$) for clockwise rotation. At the same time, the wall normal vorticity of perturbation also attains a global maximum in the same side of that of the wall normal velocity, but it has a local maximum at the opposite side of mixed layer. The absolute value of amplitude function for wall normal vorticity of perturbations attends a zero value inside the flow domain which suggests existence of more than one vorticity with different strength within the flow field. The local maximum value in the normal vorticity is appearing near to the miscible layer for anticlockwise rotation, and it is in the other side and away from the mixed layer for clockwise rotation. However, for clockwise rotation the wall normal velocity and vorticity perturbations attend a positive maxima value around the miscible layer, which may be supplying extra energy to the disturbances and thus the flow is more unstable in this case. The maximum value attended by the eigen functions alter depending on the direction of rotation. Furthermore, anticlockwise rotation is reducing vorticity-strength of the upper layer (more/less viscous for $r = 1.2/0.8$) and enhancing the vorticity-strength of lower layer (less/more viscous for $r = 1.2/0.8$). Above discussion thus far provides that paternity of the unstable disturbances depend on the viscosity contrast as well as rotational direction.

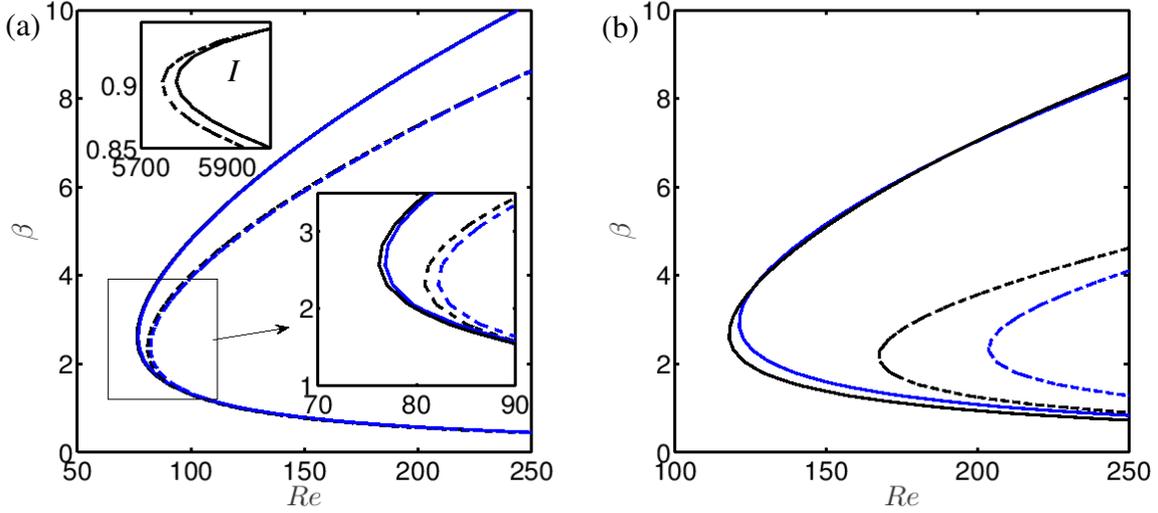

Figure 4. Neutral stability boundaries for both clockwise (dashed lines) and anticlockwise (solid lines) rotation in conjunction with more viscous fluid is near the upper wall: (a) for $r = 1.2$ and (b) for $r = 2.0$. Black lines are for miscible layer thickness $q = 0.3$ and blue curves are for $q = 0.1$. (In the inset $I$ general class of viscosity stratified flow with streamwise disturbance and without any rotation results are given)

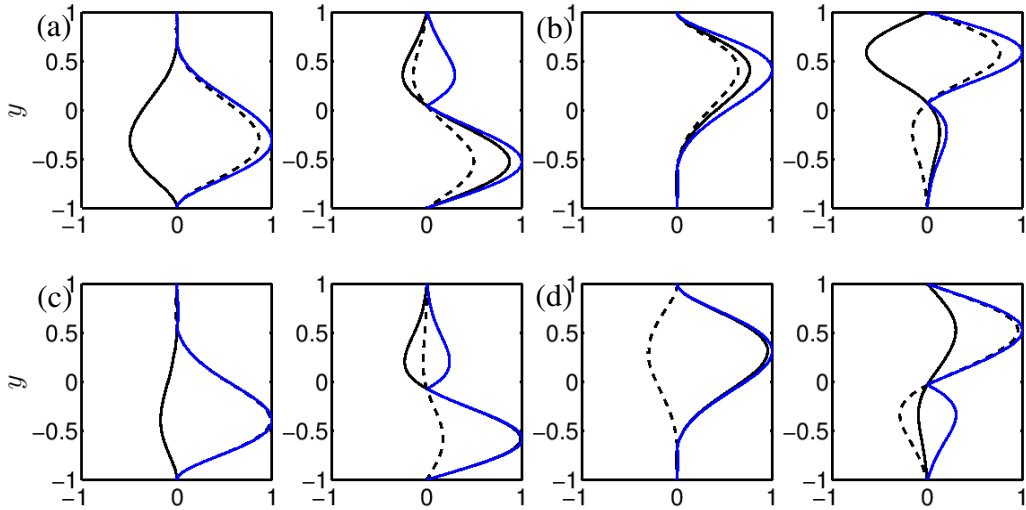

Figure 5. Typical eigen functions for different rotational directions and for two different viscosity ratios $r = 0.8$ (in (a), (b)) and $r = 1.2$ (in (c), (d)). Left pan and right pan of the subfigures represent normal velocity and vorticity, respectively. Solid black line for real part and blue line for absolute value of eigen function; dashed line are used for imaginary part. Critical values of other parameters are used as in figure are given in Table I.

We have focused on the contours of maximum growth rate $\omega_{i,\max}$ by considering all existing modes across a range of spanwise wavenumber and viscosity contrast to summarize the instability fashion at a fixed Reynolds number $(Re = 995)$. The rotation number is varied to understand the efficiency of the mixing process. Figure 6(a) and 6(b) comparison, it may thus be concluded that the mutation in the maximum growth rate for ACW and CW types of rotations. Rotation number is varied from 0.1 to 0.3, to give a practical microfluidic perspective. The growth rate for ACW rotation is high as compared to CW rotation and it increases significantly as the rotation number increases from 0.1 to 0.2. However, further increase in rotation number does not allow the growth rate to increase appreciably. We also observe that the growth rate is much high for viscosity ratio $r < 1$ and it deceases with the increase of viscosity contrast. For the sake of longer spanwise waves, the system tends to be stable for both CW and ACW rotation number.

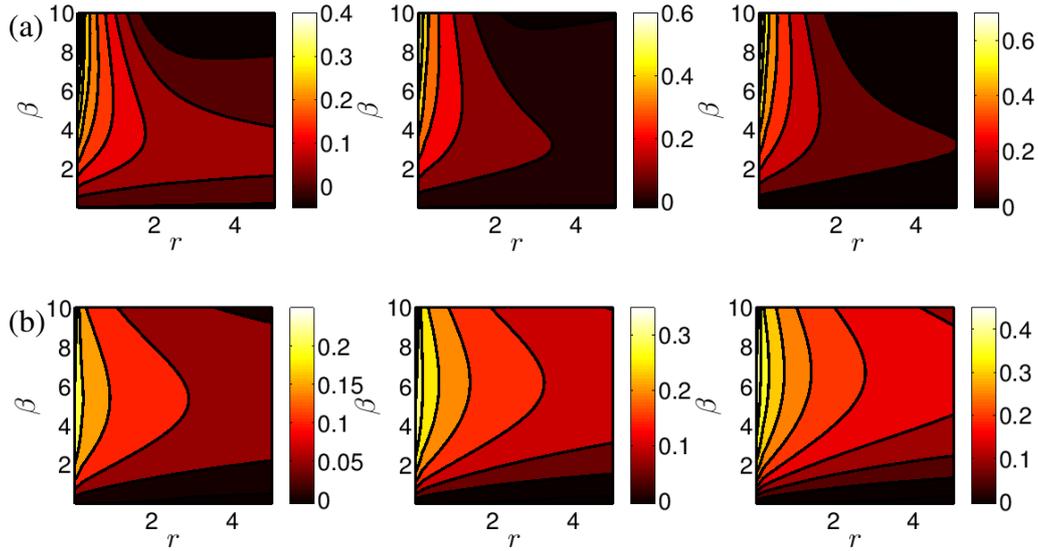

Figure 6. Contours of maximum growth rate for (a) anticlockwise and (b) clockwise rotation with three different rotation numbers $Ro = 0.1, 0.2, 0.3$. Other parameters are $\alpha = 0.0, Re = 995, p = 0, q = 0.3$.

## 3.2 Three dimensional instability

Our discussion thus far provides kind of two-dimensional (2D) (streamwise wave number is zero) instability behaviours and does not help in dealing with three-dimensional (3D) perturbations. A natural ambiguity arises when dealing with three-dimensional disturbances, concerning how viscosity contrast, rotational direction and mixed layer thickness influence the unstable modes. In Figure 7, the sets of eigen modes for 3D perturbations are plotted in the complex frequency plane for two different viscosity ratios in conjunction with different rotational directions. Existence of more than one unstable eigen mode is found for the considered set of parameters. Meanwhile, a fundamental query evolves: either the unstable modes are similar kind and sourcing from same fluid layer or their properties and origin are different. Figures 7(a), (b) confirm that some unstable modes occur due to the presence of the mixed layers (denoted by $*$symbol) but they are not the most excited. The most unstable mode with maximum growth rate is sourcing from lower/upper layer depending on the rotational direction as ACW/CW (in some cases also depends on streamwise and spanwise wave numbers). However, the utmost unstable mode is achieved with higher temporal growth for viscosity ratio $r = 0.8$, as compared to that for $r = 1.2$. The rotation number and spanwise wave number also play a crucial role in determining the behaviour of spectrum corresponding to three dimensional disturbances.

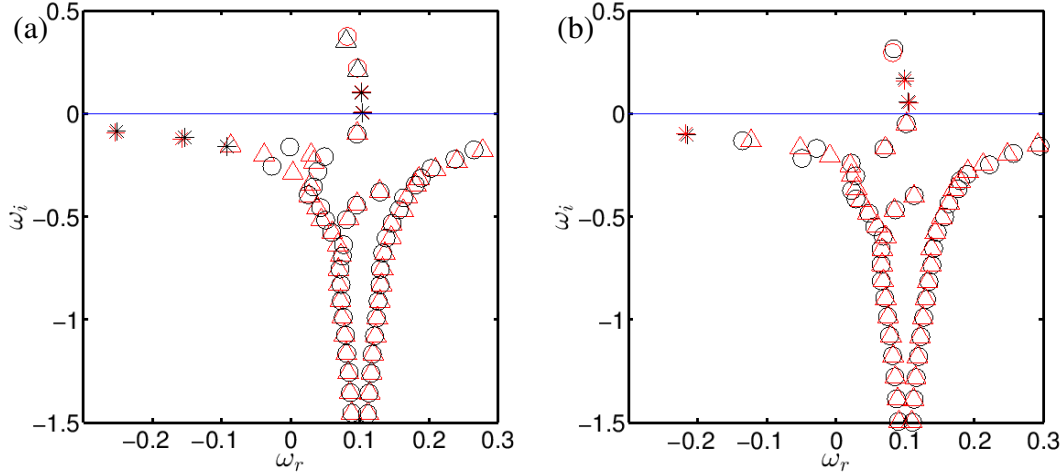

Figure 7. Eigen spectrum for $Re = 995.0, \alpha = 0.15, \beta = 1.0, Ro = 0.20$, with mixed layer thickness $q = 0.3$: in (a) $r = 0.8$ and in (b) $r = 1.2$. Different symbols are used to distinguish different modes based on their layer of sourcing. Modes, which are coming due to the presence of mixed layer are denoted by * symbol and modes from upper/lower layers are denoted by the symbol ○/△. Black and red colours suggest anticlockwise and clockwise rotation respectively.

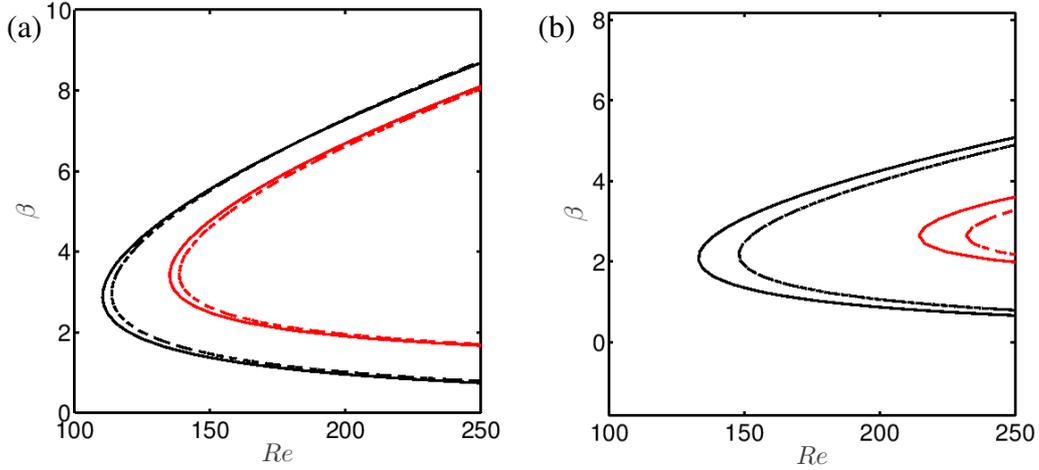

Figure 8. Marginal stability boundaries for two different viscosity ratios (a) $r = 1.2$ and (b) $r = 2.0$ for different wave numbers of streamwise disturbances $(\alpha = 0.0, 1.0)$. Solid lines are for ACW and dashed lines are for CW rotations. Critical rotation number increases with streamwise disturbances. $\alpha = 0.0$ corresponds the two-dimensional disturbance.

Properties of marginal stability boundaries with respect to three-dimensional disturbances are being discussed and compared with the two-dimensional case in Figure 8 for two different viscosity contrasts with both types of CW and ACW rotation, for rotation number $Ro = 0.2$. The Coriolis acceleration originating from system rotation gives rise to an opposite directional force, which will be normal to the walls in a rotating channel, since the basic flow is supposed to be unidirectional and parallel to the walls. This force will be directed towards the leading side of the channel depending on the direction of rotation. The basic flow will have the nearly parabolic type profile, with the largest force being slightly shifted from the centre of the channel due to the effect of viscosity stratification and coupled with an unstable 'stratification' of the Coriolis force on the leading side, and a stable one on the trailing side. For a fixed viscosity ratio, direction of the rotation plays a key role to superintend the instability nature. In this kind of system, the leading side is destabilizing and the trailing side is stabilizing one. For the case of anticlockwise rotation, higher viscous regime falls on the trailing side and less viscous regime of the flow field is on the leading side, and the reverse is true for the clockwise rotation. As seen in the figure 3, with anticlockwise rotation, the flow becomes more unstable due to the less viscous leading side, which effectively plays an important role in destabilizing the flow. In the clockwise rotation, the flow tends to be more stable due to effect of higher viscosity in the leading side. The viscosity stratification of the system is effectively suppressing the growth rate of the system as compared to single fluid rotating system, which may be useful in practical applications.

The 3D patterns of critical disturbances (corresponding to the three-dimensional perturbation) are plotted in Figures 9 and 10 to acquire additional information about the location of the stable or unstable vortices inside the flow field, where the instability may have evolved. Contours of the perturbed velocity field are displayed for viscosity ratio $r = 0.8$ and $r = 1.2$, with both ACW (Figure 9) and CW (Figure 10) rotations at a particular time. It is worth mentioning that it is quite difficult to compute the exact amplitude of eigenfunctions in the skeleton of linear stability analysis. We note that for $r < 1$, the cases corresponding to CW rotation are more unstable than ACW cases ($\omega_{i,CW} > \omega_{i,ACW}$) and reverse is true if $r > 1$. Moreover, for viscosity ratio $r = 0.8$, the liquid with higher viscosity resides in the lower layer and similarly, for viscosity ratio $r = 1.2$, the less viscous liquid remains in the lower layer.

In figures 9 and 10, the right pan shows the 2D view of the perturbation velocity contours. In left pan of the figures, we have represented three slices of the flow field at the location $z = 0$, $y = 0$ and $y = 0.3$ (near the mixed layer). For ACW rotation, in both the cases $r < 1$ and $r > 1$, primary vortices (with more strength) are formed in the lower layer. For $r = 0.8$, vortices are slightly stronger than that for $r = 1.2$ and possibly because of this, the flow system with less viscous fluid adjacent to the upper wall is more unstable. Besides, for both kinds of viscosity contrast, comparatively weaker secondary vortices are built up in the upper fluid layer of the channel. For viscosity ratio $r = 1.2$, weaker vortices are concentrated towards the mid plane of the channel, dragging the disturbances inside mixed layer towards upper half (i.e. towards higher viscous liquid) from the center line of the channel. For $r = 0.8$, the similar vortices are concentrated much closer to the upper wall of the channel and hence the penetration is less in the mixed layer.

Reversing the direction of rotation (i.e. for CW case), the location of primary vortices is shifted to the upper fluid layer (which is higher viscous fluid for $r = 1.2$ and less viscous fluid for $r = 0.8$). However, decrepit secondary vortices are present in the lower layer as well. Figure 10(a), confirms that the mixed layer is more involved in the instability mechanism for $r = 1.2$. For both the viscosity ratios considered above, the secondary vortices form near $y = -0.5$, but, strength of the vortices is comparatively lower for $r = 0.8$. We also note that for ACW case, the axis of the vortex is formed due to rotation makes an acute angle with mean flow direction, and for CW case the same makes an obtuse angle with mean flow direction. This phenomenon might have played an eminent role in the instability mechanism.

In the earlier studies on the classical single fluid Poiseuille flow with a spanwise system of rotation, the base flow profile was symmetric about the central line. For standard parabolic type profile, the largest velocity is perceptible near the central line of the channel. Since the velocity is highest at the central line, Coriolis force influenced highly near the central line, resulting an unstable flow configuration and causing instability inside the flow. However, altering the direction of rotation, the instability behaviour does not change as the mean velocity profile is balanced about the central line. On the contrary, for the case of viscosity stratified flow, the mean velocity profile is not symmetric about the central line of the channel and the skewness of the velocity profile is completely dependent on the viscosity ratio of the fluids. Probably because of the asymmetric base flow about the central line, the instability mechanism changes for viscosity stratified flow system if the direction of rotation is altered. Moreover, the base velocity gradient and modified base pressure due to presence of rotational force have also played a significant role to change the instability behaviour.

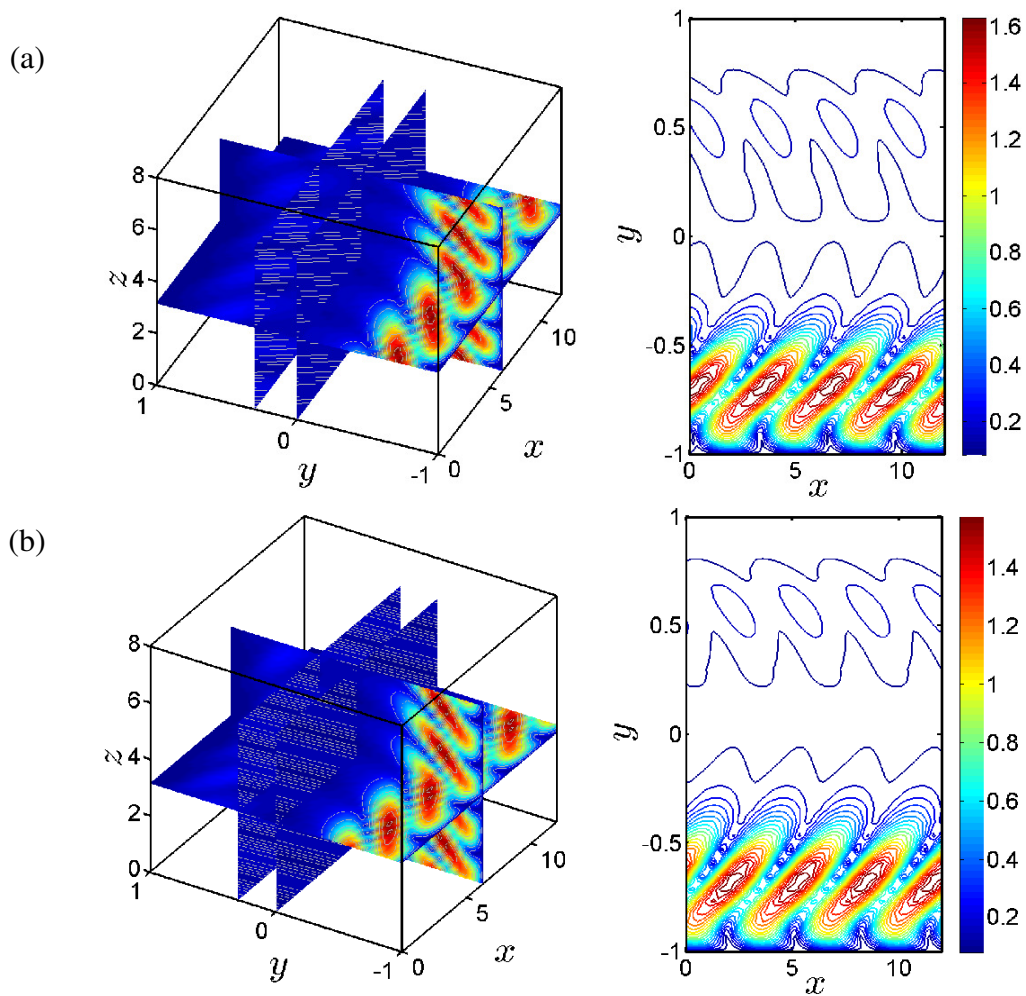

Figure 9: Iso-contours of velocity field for three dimensional disturbances with viscosity ratio (a) $r = 1.2$ and (b) $r = 0.8$ (ACW case : $Ro = 0.2, Re = 995$).

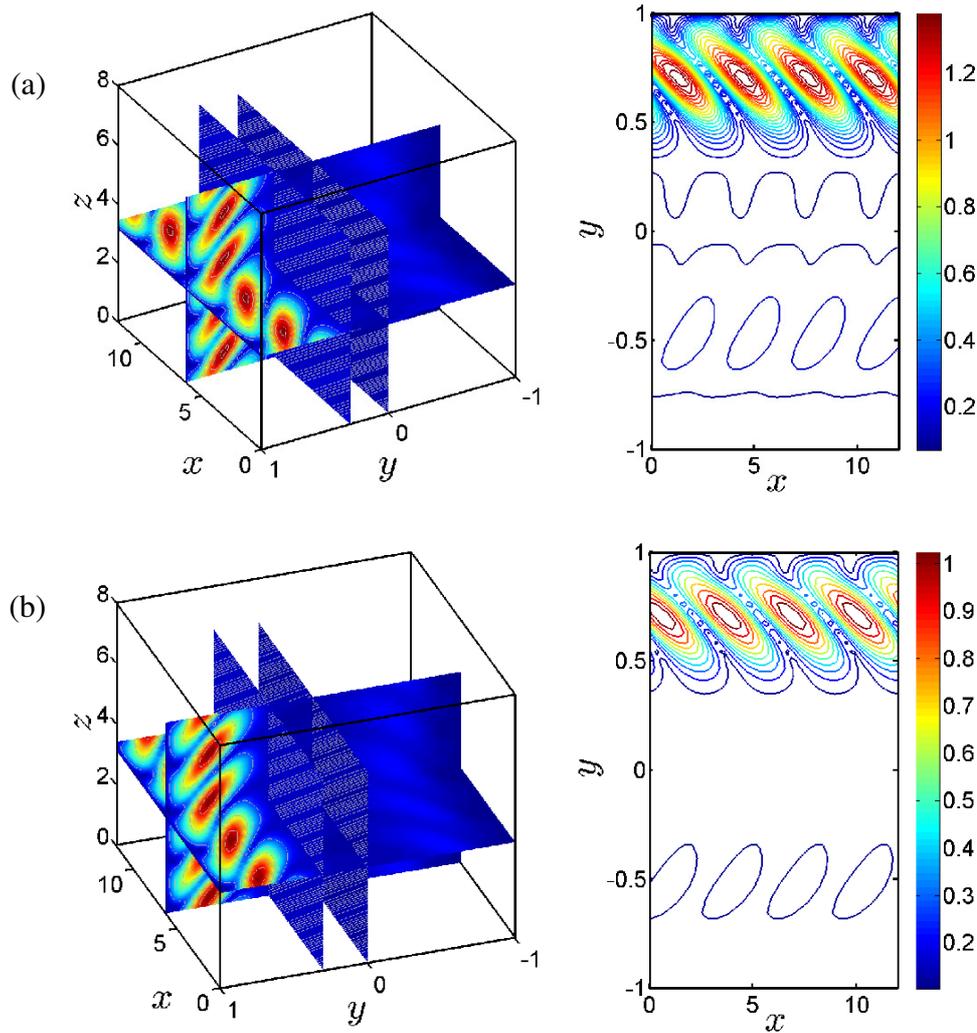

Figure 10: Iso-contours of velocity field for three dimensional disturbances with viscosity ratio (a) $r = 1.2$ and (b) $r = 0.8$ (CW case : $Ro = 0.2, Re = 995$).

In figure 11, we have plotted typical base velocity and its gradient for a viscosity stratified miscible flow with viscosity ratio $r = 2.0$. We have observed that altering the thickness of the miscible layer changes the maximum velocity, which in turn changes the range of the critical parameters (Reynolds number and Rotation number) for instability of the system due to the variation of shear rate near the mixed layer as well as centreline. It is also evident from figure 11 (b) that the velocity gradient of the base flow may be the other possible reason due to which the flow system with thicker miscible layer is more unstable. Furthermore, this idea is a direct consequence of Bradshow's instability criterion (Bradshaw, 1969) developed for a rotating flow system.

For the case of viscosity stratified miscible flow the existing literature (Ghosh et al., 2014; Govindarajan, 2004; Malik and Hooper, 2005), indicates that the thinner mixed layer is more unstable for streamwise disturbances. On the contrary to above stated fact as also discussed earlier our numerical findings shows that viscosity stratified flow with a spanwise rotation can be more unstable for thicker mixed layer. To understand underlining physics behind this contradiction, we have tried to show the energy variation of the perturbed system represented by $U'uv$ ($u, v$ are the perturbation velocity components), drown from base flow to transfer disturbed flow. The same is plotted as a function of $y$ for two different miscible layer thickness ($q = 0.1$ and $0.3$) with both ACW and CW rotation as shown in figure 12. We consider the case $r > 1$ with two different viscosity ratio $r = 1.2$ in (a) and $2.0$ in (b). For both the cases, we see the thick miscible layer draws energy more efficiently from the mean flow, which can cause the flow to be more unstable for spanwise system of rotation. Altering the rotation direction the qualitative behavior has not change. Moreover, to get the precise physical mechanism further critical instigation is needed.

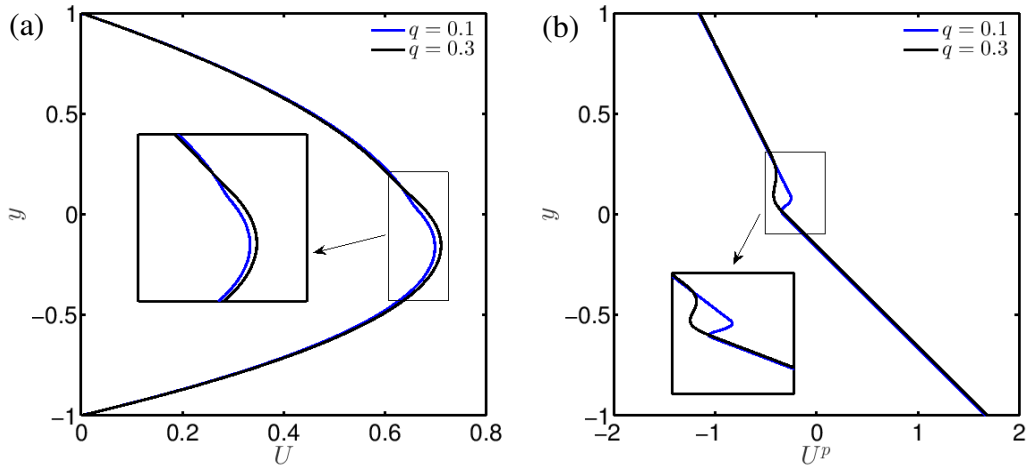

Figure 11. (a) Base velocity profile and (b) gradient of the base flow profile for viscosity ratio $r = 2.0$ and two different mixed layer thickness $q = 0.1, 0.3$.

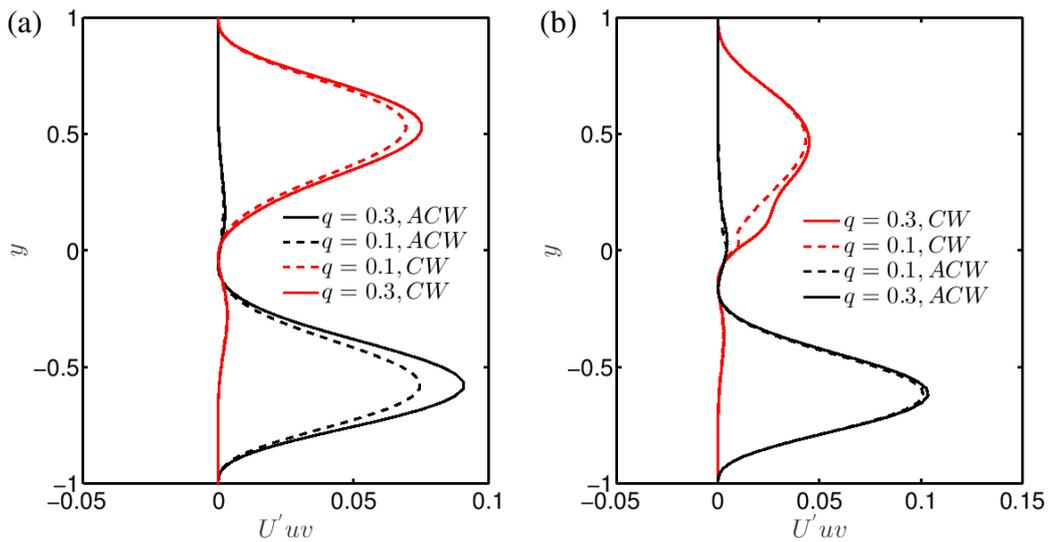

Figure 12: Variation of the energy transfer term for viscosity ratio (a) $r = 1.2$ and for (b) $2.0$. The other parameters are the critical values as mentioned in Table I.

## 4. Conclusions

Present study discussed the effects of the Coriolis forces towards influencing the stability regimes in rotating laminar miscible viscosity stratified channel flow. The linear instability behaviour and possible mechanisms are argued with reference to streamwise and spanwise disturbances (in the context of two and three dimensional configuration). The influence of viscosity ratio and thin diffusive layer called as mixed layer between two different fluids is taken into consideration in this investigation. We have presented the significance of mixed layer thickness ($q$) and the position of the mixed layer ($p$). Modal stability analysis of the flow ascertains the variation of instability mechanism for the viscosity stratified flow depending on rotational direction. Numerical results pinpoint the ranges of various critical parameters, namely rotation number and Reynolds number for clockwise and anticlockwise rotation. Moreover, clockwise/anticlockwise rotation gives necessary effort to make the flow more unstable as compared to that of the anticlockwise/clockwise (ACW/CW) rotation for viscosity ratio $r < 1 / r > 1$, respectively.

In contrast to general non-rotating class of viscosity stratified flows, our analysis indicates that with spanwise disturbances, the rotating stratified flow with thicker miscible layer is more unstable than that of the flow with thinner mixed layer. This is the most important and novel finding of the current work. Assessment of maximum growth rate over a range of spanwise disturbance has been conducted for different viscosity ratios, by varying the rotation number, to understand the efficiency of the mixing process. It is found that the growth rate for ACW rotation is eminent as compared to CW rotation and it increases significantly when the rotation number increases. We have also tried to understand qualitatively the three-dimensional perturbed flow field and how clockwise and anticlockwise rotation can affect the flow vortices to control the mixing process. A critical interaction between the vortices present near and away from the mixed layer is seen and it explains the alternation of instability due to ACW & CW rotation. An eigen-spectrum analysis reveals that for three-dimensional disturbances, unstable modes sourcing from different layers can play a monumental role in the temporal instability depending on viscosity ratio ($r$) of the fluids. Results displayed the existence several unstable modes, some of which are originating from mixed layer. Appearance of the unstable modes from different layers relies on rotational direction and spanwise disturbances. However, the system with less viscous fluid adjacent to the upper wall is more unstable than the opposite configuration and for this configuration most excited mode belongs to the lower/upper layer depending on the rotational direction as ACW/CW. The possible reasons for the changes in instability behaviour depending on viscosity ratio, rotational direction, mixed layer thickness etc. are the outcome of the change in maximum velocity, relative flow rate and skewness of base velocity profile, velocity gradient as well as the shear rate by virtue of the viscosity stratification. The dispersion of energy transport term $U'uv$ as a function of $y$ suggests that the flow with the thick miscible layer draws energy more efficiently from the mean flow, which may be implicitly causing the flow to be more unstable in the presence spanwise system rotation.